# Update hydrological states or meteorological forcings? Comparing data assimilation methods for differentiable hydrologic models

Amirmoez Jamaat[1], Yalan Song[1*], Farshid Rahmani[1], Jiangtao Liu[1], Kathryn Lawson[1], Chaopeng Shen[1*]

[1] Department of Civil and Environmental Engineering, The Pennsylvania State University, University Park, PA 16802

* Corresponding authors: Chaopeng Shen, cshen@engr.psu.edu; Yalan Song, YXS275@psu.edu

## Abstract

Data assimilation (DA) enables hydrologic models to update their internal states using near-real-time observations for more accurate forecasts. With deep neural networks like long short-term memory (LSTM), using either lagged observations as inputs (called "data integration") or variational DA has shown success in improving forecasts. However, it is unclear which methods are performant or optimal for physics-informed machine learning ("differentiable") models, which represent only a small amount of physically-meaningful states while using deep networks to supply parameters or missing processes. Here we developed variational DA methods for differentiable models, including optimizing adjusters for just precipitation data, just model internal hydrological states, or both. Our results demonstrated that differentiable streamflow models using the CAMELS dataset can benefit strongly and equivalently from variational DA as LSTM, with one-day lead time median Nash-Sutcliffe efficiency (NSE) elevated from 0.75 to 0.82. The resulting forecast matched or outperformed LSTM with DA in the eastern, northwestern, and central Great Plains regions of the conterminous United States. Both precipitation and state adjusters were needed to achieve these results, with the latter being substantially more effective on its own, and the former adding moderate benefits for high flows. Our DA framework does not need systematic training data and could serve as a practical DA scheme for whole river networks.

## Highlights

1. Variational data assimilation (DA) can boost differentiable models' performance.
2. Differentiable models with DA outperformed LSTM-DA in most US except the West.
3. Both precipitation and state adjusters were needed to achieve the best result.

# 1. Introduction

Accurate streamflow forecasts are crucial for disaster mitigation and water resources management (Merz et al., 2021). In recent years, there have been rising flood risks around the world as the water cycle intensifies (IPCC, 2021). Hydrologic models which can simulate streamflow responses to atmospheric inputs (precipitation, temperature, etc.) serve as the basis for issuing flood warnings. They also provide vital information for aiding in drought planning (Alfieri et al., 2013; McEnery et al., 2005; Van Loon, 2015). However, because of model structural inadequacies, parametric uncertainty, and unpreventable meteorological forcing data errors, significant errors can linger and accumulate in the model's hydrological states which lower their forecast accuracy.

Several ways of incorporating near-real-time streamflow observations have been shown to improve operational hydrologic forecasts (Ercolani & Castelli, 2017; Kurtz et al., 2017; Leach & Coulibaly, 2019; Liu et al., 2012). The standard approach is data assimilation (DA), which updates the model's state variables by minimizing the difference between historical predictions and the actual observations. DA approaches including ensemble Kalman filter (EnKF), particle filter (PF), and variational DA have been used extensively in hydrology (Camporese et al., 2009; Ercolani & Castelli, 2017; Liu et al., 2012; Mazrooei et al., 2019; McLaughlin, 1995; Reichle, 2008). The first two approaches build an ensemble of model simulations to represent the uncertainty and the potentially nonlinear relationship between model states and the observable variable, then use this relationship (encoded in the covariance matrix) to guide the adjustment of states so that the forecast is improved (Moradkhani et al., 2005; Pasetto et al., 2012). As an alternative, variational DA relies on the calculation of gradients of variable simulations (which correspond to observations) with respect to the adjustable states, parameters, or inputs, and uses such gradients to optimize the match with observations. Variational DA is "training-free" --- it solves an optimization problem, and is thus applicable to new situations without extensive historical data. It is widely applied in both weather forecasting (Bouttier & Kelly, 2001) and hydrologic modeling (Cheng et al., 2021; Ercolani & Castelli, 2017; Hernández & Liang, 2018). However, there are often a large number of adjustable variables and not enough constraints, leading to substantial uncertainty in the optimal adjustments. Also, for complex models, the significant effort of building the software infrastructure needed to calculate the gradients can be daunting.

Currently, deep learning (DL) techniques for time-series analysis have become powerful tools for streamflow predictions (Camps-Valls et al., 2021; Shen, 2018). Among the available algorithms, Long Short-Term Memory (LSTM) networks have proven to be particularly successful as forward (prediction) models for streamflow simulation (Feng et al., 2020; Kratzert et al., 2019). These models are capable of capturing temporal dependencies and making correlations between datasets. Taking advantage of the capacity of deep networks like LSTM to automatically learn the best way of using inputs, a natural way for them to ingest recent streamflow data is data integration (DI), which is similar to the autoregression used in statistical hydrologic models (Fang & Shen, 2020; Feng et al., 2020; Moshe et al., 2020) directly include recent streamflow observations along with other forcings and geographic attributes as inputs to predict future streamflow. Essentially, DI lets the neural network automatically learn how to best utilize the recent observations to improve future forecasts. Using LSTM with DI, Feng et al. (2020) improved streamflow performance from a median NSE of 0.73 to 0.86 on the CAMELS dataset of 671 basins and one source of meteorological forcing data. Due to its simple implementation and performance, this approach has become popular in deep learning (DL) models (Khoshkalam et al., 2023; Mangukiya et al., 2023; Yao et al., 2023). However, this type of data integration depends on the availability of sufficient systematic training data so that the DL model can learn the relevant relationships - otherwise the resulting models may degrade in effectiveness when applied to untrained sites.

With the development of deep learning models, there are alternative methods for data assimilation, such as using neural networks to directly map the target variable to states variable (Boucher et al., 2020) or using automatic differentiation (AD) for gradient tracking in variational DA (Nearing et al., 2022). AD in languages like PyTorch, TensorFlow, Julia, and JAX significantly alleviated the challenges of calculating gradients, which previously required complex adjoint models (Innes et al., 2019; Shen et al., 2023). These platforms record the computational graph during the forward model run (prediction task) and can easily (although not computationally trivially) provide the gradients of outputs with respect to inputs. These platforms became marvelously efficient at gradient calculation on Graphical Processing Units (GPUs). Leveraging the gradients obtained from AD, Nearing et al. (2022) implemented variational DA with LSTM by optimizing the objective function of the historical streamflow to update its cell states and predict the future streamflow with the updated states, achieving a median NSE of 0.86 on the CAMELS dataset when multiple forcing inputs were employed. The gradients were used as a means of optimizing the variational DA problem, rather than training neural network weights. However, the performance is lower than the DI methods. Furthermore, LSTM and other pure DL models have limitations including interpretability and spatial extrapolation (Feng et al., 2023, 2024). Additionally, they cannot output variables that were not the training targets (training targets are typically limited to observable variables with sufficient available observation data) and may not represent extreme events well (Song, Sawadekar, et al., 2024).

Recently, differentiable models, a genre of physics-informed machine learning, have demonstrated state-of-the-art accuracy paired with physical interpretability (Feng et al., 2022; Shen et al., 2023). In contrast to physics-informed neural networks (PINN), which are DL models trained as simple surrogates for process-based models, differentiable models directly embed the physics inside the model system, training connected neural networks along with the equations at the same time (in an "end-to-end" fashion). They inherit the learning capability, large-domain generalizability (Fang et al., 2022), and highly efficient training on GPUs (Jeon et al., 2021; Wang et al., 2022) characteristic of DL, while also maintaining process clarity and the ability to predict variables beyond the training targets. (Feng et al., 2022) developed a differentiable Hydrologiska Byråns Vattenbalansavdelning ($\delta$HBV) model for streamflow prediction, which incorporated an LSTM network to calculate and provide static or dynamic parameters to the physical equations from HBV. Unlike a standalone LSTM, "ground truth" parameters were unnecessary, since the re-implemented HBV equations were also differentiable and the end-to-end training capability enabled the LSTM to be constrained by comparing the resulting streamflow predictions to actual observations. Later work achieved comparable accuracy to pure LSTM streamflow models (Song, Knoben, et al., 2024). Due to the physical constraints enforced by the connected physical model "backbone", differentiable models have demonstrated better spatial extrapolation in streamflow prediction for ungauged regions (Feng et al., 2023) and unseen extreme events (Song, Sawadekar, et al., 2024) than the previous state-of-the-art LSTMs. (Song, Bindas, et al., 2024) further developed a high-resolution differentiable model and a seamless streamflow product for ~200,000 river reaches (with a median length of ~7 km) across the conterminous United States (CONUS) for national-scale operational flood prediction. Although DA has not yet been implemented for differentiable models, it is reasonable to expect that it will lead to some improvements in model performance. However, since the current differentiable models have rather small numbers of physically-meaningful state variables compared to LSTM's large hidden sizes, it remains unclear if variational DA can benefit differentiable models as much as it benefited LSTM. It is further unclear which types of errors in differentiable models can be reduced through its application.

There are multiple ways in which DA can be applied to differentiable models: the hydrological state variables and the input forcings can each be adjusted, and benefits from either or both options are not guaranteed. Hydrological state errors are attributed to issues with the model's mechanisms and result in an incorrect distribution of mass --- correction via DA could resolve lingering (long-term) errors in the system, but could also introduce abrupt (short-term) changes in the streamflow predictions. Errors with meteorological forcings result from the processes used to derive this information; correction via DA offers the opportunity to consistently revise the hydrological states across the entire simulation (i.e., enable short-term corrections), but may not account for long-term accumulated errors. An assessment of the effectiveness of these options has not been done.

In this work, we implemented variational DA on the differentiable HBV model developed by (Feng et al., 2022) and further improved by Song et al. (2024), called δHBV1.1p-CAMELS-hydroDL (δHBV for short). We used DA for three distinct purposes: (1) correcting biases in the precipitation data by adjusting the dynamic weights assigned to precipitation days, (2) reducing long-term errors by updating the states in the HBV model, and (3) simultaneously reducing errors in both the precipitation and state variables. We sought answers to the following questions:

1. For streamflow forecasting, can differentiable models with limited physical states benefit as much from variational DA as LSTM?
2. Does the structure of a differentiable model like δHBV impose any limitations on DA? If so, do these limitations prevent δHBV from matching the performance of LSTM with DA?
3. What is the most effective combination of model hydrological state and/or precipitation adjusters for variational DA with differentiable models?

## 2. Methods

*2.1. Datasets*

We employed the Catchment Attributes and Meteorology for Large-Sample Studies (CAMELS) dataset (Addor et al., 2017; Newman et al., 2015) to inform and evaluate the streamflow predictions of δHBV and its variants with different DA methods. Model inputs included the basin-averaged meteorological forcing data from Daymet (Thornton et al., 1997) as included in CAMELS, as well as geographic attributes from CAMELS (see Table A1 for details). We utilized 531 basins from a total of 671 CAMELS watersheds as selected by Newman et al. (2017), who removed the basins with large discrepancies in area calculations across different methods, as well as those with catchment areas larger than 2000 km$^2$.

*2.2. Models*

To summarize, we used two "backbone" models — the differentiable HBV model (δHBV) and LSTM — and compared the effectiveness of several variational DA approaches for these models on the CAMELS dataset. For LSTM, DA is used to adjust its hidden states (The structures of LSTM and DA applied to LSTM are adapted from Kratzert et al. (2019) and Nearing et al. (2022)). In contrast, DA for δHBV can adjust the forcing (precipitation in this work) input, the physical states of HBV, or both. Additionally, we employed data integration (DI) for LSTM as a highly competitive benchmark from Feng et al. (2020). Here, we provide a brief introduction to these methods and focus primarily on the three DA algorithms developed for δHBV.

*2.2.1. Differentiable HBV model*

The original HBV model is a conceptual rainfall-runoff model that can simulate snow accumulation and melt, soil moisture, evapotranspiration, and runoff generation from different layers of soil storages. The differentiable model, δHBV1.1p (Feng et al., 2022; Song, Sawadekar, et al., 2024), is implemented on the PyTorch platform to enable efficient gradient tracking with automatic differentiation throughout the entire model, from the physical model components to the neural network, so the framework can be trained on large datasets. Since this model has been thoroughly documented and benchmarked against LSTM in the literature, we briefly summarize it as follows (additional information of HBV model is available in the Appendix C):

$$\theta_{b,s}, \theta_{b,d}^t = \text{LSTM}(x_b^t, A_b) \tag{1}$$

$$Q_b^t, S_b^t = \text{HBV}(\theta_{b,s}, \theta_{b,d}^t, x_b^t, S_b^{t-1}) \tag{2}$$

Here, $\theta_{b,s}$ and $\theta_{b,d}^t$ represent the groups of static (subscript $s$) and dynamic (subscript $d$, with superscript time step $t$) parameters for each basin, $b$. In δHBV $x_b^t$ represents the time-dependent meteorological forcings (precipitation, mean temperature, and potential evapotranspiration calculated using the method from Hargreaves et al. (1994)), and $A_b$ denotes the static geographical attributes (listed in Table A1 in Appendix). $S_b^{t-1}$ are the hydrological states in δHBV from the previous time step and initially set to zeros, which include soil moisture (SM, the water content in the soil), snowpack (SWE, or snow water equivalent - the accumulated water present as snow), meltwater (the liquid water present in the snowpack), upper zone storage (SUZ, the upper subsurface water storage that affects quick flow components like surface runoff), and lower zone storage (SLZ, the lower subsurface water storage that affects the release of baseflow). $Q_b^t$ and $S_b^t$ are the simulated streamflow and states for basin $b$ at time $t$, respectively. LSTM is the network used to determine the parameters, while HBV represents the physical equations (Appendix C).

During both the model training and data assimilation phases, we employed the following normalized square-error loss function:

$$\text{Loss}(Q_b^{t:T}, Q_b^{*\,t:T}) = \frac{1}{B}\sum_{b=1}^{B}\sum_{t=1}^{T}\frac{(Q_b^t - Q_b^{*\,t})^2}{(\sigma_b^t + \varepsilon)^2} \tag{3}$$

where $B$ represents the total number count of basins in each batch, and $T$ denotes the time length of the simulated streamflow time series. $Q_b^t$ and $Q_b^{*\,t}$ correspond to the simulated and observed streamflow values, respectively, for basin $b$ and time $t$. $B$ represents the total number of basins in the batch. $\sigma_b$ is the standard deviation of the observed streamflow for basin $b$, which is included to prevent a disproportionate influence from large or more hydrologically active basins within the loss calculation. $\varepsilon$ is a small constant introduced to ensure the denominator does not become zero. Before starting the DA procedure, the δHBV base model is trained on the forcings, basin attributes, and streamflow data with the configurations in the following section. After training, its physical states and parameters over the forecasting timespan are saved and frozen (prevented from changing, unless otherwise specified like for physical states in state adjustment DA).

*2.2.2. DA with precipitation adjustment*

In the precipitation adjustment version of δHBV with DA (called δHBV-DA-P), we utilized a multiplier adjuster for short-term precipitation inputs in the assimilation window to minimize the loss function of δHBV-DA-P and enhance its streamflow forecast in the next evaluation day. For the streamflow forecast at time $t$+1, we adjusted the precipitation data at day $t−a$ in the DA window (Figure 1) with a parameter, $k$, which is determined by minimizing the loss calculated from day $t−w+1$ to day $t$ ($w$ is the length of the assimilation window). We tested assimilation window lengths of 1, 3, 5, 10, and 20 days across all basins, with $w = 5$ showing the highest performance in terms of streamflow prediction accuracy. We also tested precipitation adjustment at different days ($a$) within the 5-day assimilation window, including single-day adjustments ($a = $ 1, 2, 3, 4, or 5) and multiple-day spans. In our case, we found that $a = 2$ provided the highest performance improvement for streamflow forecasting. To consider the nonlinearity of the systematic errors in the precipitation data, we tested different types of functions of $k$ as the precipitation amplifier, including linear ($mk + n$), exponential ($e^k$), and polynomial functions ($k^2$, $k^4$, $k^6$). We chose and employed $k^4$ as a scalar to adjust the precipitation at $t−a$ within the data assimilation window, as it provided the best performance in enhancing the streamflow simulation.

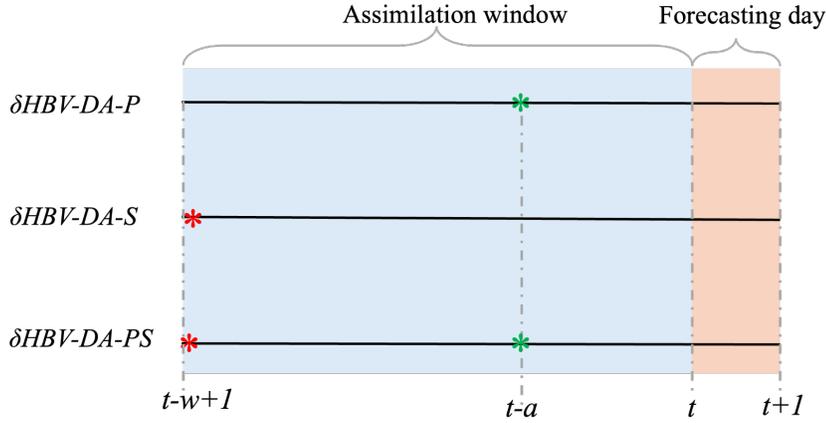

*Figure 1: Schematic view of DA methods for δHBV in an assimilation window of length w. Streamflow observation from day t−w+1 to day t are assimilated for a prediction at day t+1. δHBV-DA-S adjusts hydrological states (red star) at day t−w, δHBV-DA-P adjusts precipitation (green star) at day t−a, and δHBV-DA-PS has adjustments of both hydrological states (at day t−w+1; red star) and precipitation (at day t−a; green star).*

The procedures for DA precipitation adjustment are as follows:

1. For each prediction (at day $t+1$), the preceding $w = 5$ days are the DA window (day $t-w+1$ through day $t$). The precipitation ($p$) for each basin ($b$) at day $t-a$ (for this work, $a = 2$) is adjusted to $p'$ by multiplying $k^4$ ($p$ is one component of $x$, the meteorological forcings):

$$p'^{(t-a)}_b = k^4_b p^{(t-a)}_b \qquad (4)$$

2. Simulations for each day within the DA window (day $t-w+1$ through day $t$) are run from day $t-w+1$ using the adjusted precipitation, along with the saved physical states and parameters, to calculate the adjusted streamflow within the DA window:

$$Q'^{(t-w+1):t}_b = HBV\left(\theta_{b,s}, \theta^{(t-w+1):t}_{b,d}, x^{(t-w+1):t}_b, p'^{(t-a)}_b, S^{(t-w+1)}_b\right) \quad (5)$$

3. Thanks to the differentiability of δHBV, the gradient of the loss with respect to $k$ can be automatically calculated using automatic differentiation. The AdaDelta optimizer based on gradient descent is used to optimize the loss function (same as Eq. 3):

$$k'_b = argmin_{k_b}\left(\text{Loss}\left(Q'^{(t-w+1):t}_b, Q^{*\,(t-w+1):t}_b\right)\right) \quad (6)$$

If the loss function within the window is decreased after DA, we accept the adjustment and use the updated precipitation. Otherwise, we use the original precipitation. The adjustment is not applicable on dry days.

4. For each day, we perform steps 1 to 3 on a batch of basins to take advantage of GPU parallelization. The loss function is evaluated for the DA window of all basins in the batch, and the optimizer simultaneously updates $k$ for all basins in the batch based on a single loss value. It should be noted that this is different from adjusting neural network weights, as $k$ is specific to each basin in the assimilation window.

5. After adjustment, the model is run "forward" again using the saved physical states and parameters but the updated precipitation (if accepted in step 3), to the forecasting day (the next day after the DA window, day $t$+1).

$$Q'^{(t-w+1):t+1}_b = HBV\left(\theta_{b,s}, \theta^{(t-w+1):t+1}_{b,d}, x^{(t-w+1):t+1}_b, p'^{t-a}_b, S^{t-w+1}_b\right) \quad (7)$$

6. We take $Q'^{t+1}_b$ as the new prediction at day $t$+1. Steps 1 through 5 are repeated for each day within the forecasting period.

*2.2.3. DA with state adjustment*

The state adjustment version of δHBV with DA (called δHBV-DA-S), is intended to enhance streamflow forecasts by correcting the model's internal hydrological states. By adjusting the initial states at the beginning of DA window (day *t−w+1,* Figure 1), discrepancies between simulated and observed streamflow can be minimized.

The procedures for DA state adjustment are as follows:

1. For each forecasting day (at *t+1*), the preceding *w* = 5 days are the DA window. We conduct the streamflow simulation within the DA window using (adjusted or saved) states, $S'^{(t-w+1)}$, to calculate the (adjusted) streamflow:

$$Q'^{(t-w+1):t}_b = \text{HBV}\left(\theta_{b,s}, \theta^{(t-w+1):t}_{b,d}, x^{(t-w+1):t}_b, S'^{(t-w+1)}_b\right) \quad (9)$$

2. $S'^{(t-w+1)}$ is optimized using gradient descent to minimize the loss function (same to Eq. 3) of the DA window.

$$S'^{(t-w+1)}_b = argmin_{S^{(t-w+1)}_b}\left(\text{Loss}\left(Q'^{(t-w+1):t}_b, Q^{*\,(t-w+1):t}_b\right)\right) \quad (10)$$

If the loss function is decreased after DA, the state updates are accepted. Otherwise, the original states before optimization are used. Similarly, the model evaluates the loss function over the DA window for each batch of basins, and simultaneously adjusts their hydrological state variables $S$ based on the single aggregated loss value.

3. We perform step 1 to 2 for one DA window. After the state adjustment, the model is run forward using the adjusted physical states to obtain streamflow prediction for day t+1:

$$Q'^{(t-w+1):t+1}_b = \text{HBV}\left(\theta_{b,s}, \theta^{(t-w+1):t+1}_{b,d}, x^{(t-w+1):t+1}_b, S'^{t-w+1}_b\right) \quad (11)$$

4. We take $Q'^{t+1}_b$ as the new prediction at day $t+1$. Steps 1 to 3 are repeated for each day in the forecasting period.

*2.2.4. DA with state and precipitation adjustment*

For combined precipitation and state adjustment within the differentiable HBV model (δHBV-DA-PS), we aimed to adjust both the short-term precipitation inputs and the internal hydrological states simultaneously from historical days to minimize the loss of the DA windows (Figure 1). The procedures are exactly as described above, only that both of them are employed.

*2.2.4 Data assimilation and data integration for LSTM*

LSTM in δHBV (Eq. 1) is only used for generating parameters and trained indirectly with the output from the HBV model. In our study, we used three additional LSTM models for direct streamflow prediction as benchmarks, which are pure machine learning models without physics incorporated.

The first is a standalone LSTM for streamflow that is given on atmospheric forcings ($f^t$) and static site attributes ($A_b$) to predict streamflow ($Q^t_b$).

$$Q^{1:t}_b = \text{LSTM}(x^{1:t}_b, A_b) \tag{12}$$

The structure of LSTM for streamflow (detailed in Appendix B) is similar to the LSTM structure used in δHBV. Unlike the LSTM in Eq. 1, which is used for HBV parameterization with only three forcing variables, the LSTM for streamflow prediction utilized all available forcing variables, $x$, from Daymet, as provided by the CAMELS dataset, including precipitation, solar radiation, maximum and minimum temperatures, and vapor pressure. The attributes, A, include topography, climate, soil texture, land cover, and geology, similar to those in Table A1, but with some non-numeric attributes removed. Full list can be found in Kratzert et al. (2019).

The second LSTM model is the LSTM-DI model developed by Feng et al. (2020), which uses the same forcing variables in Eq. 1 and attributes from Table A1. This model integrates most recent (1-day-lagged) streamflow observations—the same approach as the autoregressive method described in Nearing et al. (2022)—to dynamically adjust both states (cell and hidden states) and weights of LSTM, thereby minimizing the accumulation of compounding errors.

$$Q_b^{2:t} = LSTM(x_b^{2:t}, A_b, Q_b^{1:t-1}) \qquad (13)$$

The third LSTM model used data assimilation (LSTM-DA) following the strategy in Nearing et al. (2022), where the model's cell and hidden states are updated based on loss over a window period. In each five-day window (w = 5, time t−w+1 to time t), the model adjusts LSTM's states when the loss of the streamflow simulation in the DA window decreases (Eq. 14 and Eq. 15). They repeatedly update states with 20 windows to assimilate 100-day observations, then use the updated states to predict streamflow next to the last DA window at time step t+1 (Eq. 16):

$$Q_b^{(t-w+1):t} = \text{LSTM}\left(x_b^{(t-w+1):t}, A_b, c_b, h_b\right) \qquad (14)$$

$$c'_b, h'_b = argmin_{c_b, h_b}\left(\text{Loss}\left(Q'^{(t-w+1):t}_b, Q_b^{*\,(t-w+1):t}\right)\right) \qquad (15)$$

$$Q'^{(t-w+1):t+1}_b = \text{LSTM}\left(x_b^{(t-w+1):(t+1)}, A_b, c'_b, h'_b\right) \qquad (16)$$

where $c$ and $h$ are cell and hidden states, respectively, which are updated by DA to $c'$ and $h'$. $x$ and $A$ are forcing and attributes same to Eq. 12. Full list can be found in Nearing et al. (2022).

### 2.3. Model training and evaluation

The background $\delta$HBV and LSTM models were both trained on 531 CAMELS basins, with streamflow observational data from 1999/10/01 to 2008/09/30 over 100 epochs, with a batch size of 100. LSTM in $\delta$HBV for parameterization and LSTM directly used for streamflow both used a hidden size of 256 and a sequence length of 365 days. We first tested the model from 1989/10/01 to 1999/09/30 to save HBV parameters and states and conducted data assimilation in the same time span. Same training and testing configuration were used in Kratzert et al. (2019) and Nearing et al. (2022)).

The models were evaluated using multiple metrics. Nash-Sutcliffe Efficiency (NSE, (Nash & Sutcliffe, 1970)) measures how well the predicted flows match observed data, while Kling-Gupta Efficiency (KGE, (Gupta et al., 2009)) assesses overall agreement by considering correlation, bias, and variability. For both of these metrics, a value of 1 would indicate perfect performance. We also calculated the root-mean-square error (RMSE) for low and high flows — specifically, RMSE of the bottom 30% flow and of the top 2% flow — to evaluate accuracy of different flow magnitudes. For these, values closest to zero are desired. Additionally, we examined the annual relative peak errors (%) to assess how well the models predict annual peaks.

## 3. Results and discussion

In the following, we benchmarked the variational DA schemes using the state adjuster ($\delta$HBV-DA-S), precipitation adjuster ($\delta$HBV-DA-P), and both ($\delta$HBV-DA-PS), against purely data-driven LSTM, LSTM with data integration (LSTM-DI) and LSTM with variational data assimilation (LSTM-DA). We evaluated the relative contributions of the precipitation and state adjusters, then compared their performance in different regions of the USA, and examined the actions taken by the DA algorithm.

### 3.1. Comparing DA schemes

The state adjuster ($\delta$HBV-DA-S) was the main effective DA mechanism which reduced both the high- and low-flow RMSE, while the precipitation adjuster introduced a modest improvement in high-flow RMSE, making the combined DA ($\delta$HBV-DA-PS) the best DA scheme (median NSE (KGE) of 0.82 (0.85); see Table 1). Individually, δHBV-DA-S substantially reduced both high-flow RMSE (by 0.67 mm/day; from 3.36 mm/day of the δHBV model to 2.69 mm/day) and low-flow RMSE (by 0.05 mm/day; from 0.06 mm/day to 0.01 mm/day), resulting in a median NSE (KGE) of 0.82 (0.84). In contrast, the precipitation adjuster alone was only moderately effective, reducing high-flow RMSE by 0.21 mm/day (to 3.15 mm/day) and low-flow RMSE by 0.01 mm/day (to 0.05 mm/day). The near-real-time (lagged by two days) precipitation adjuster's lackluster effectiveness could result from its restriction to short-term impacts on direct runoff, since it cannot correct errors that happened a long time ago, e.g., with long-term water storage and snow accumulation processes. When we added the precipitation adjuster to δHBV-DA-S, it did not further improve low flow, but did have a moderate benefit for the high flow beyond the state adjuster. This observation suggests precipitation error matters more for quick flow processes that cannot be addressed by state adjustment alone.

Table 1: Comparative performance metrics for δHBV before and after data assimilation, benchmarked with Daymet forcings for 531 CAMELS basins. Values are the median for all basins.

| Model | NSE | KGE | Low Flow RMSE (mm/day) | High Flow RMSE (mm/day) |
|---|---|---|---|---|
| δHBV | 0.75 | 0.754 | 0.06 | 3.36 |
| δHBV-DA-P | 0.74 | 0.800 | 0.05 | 3.15 |
| δHBV-DA-S | 0.82 | 0.841 | 0.01 | 2.69 |
| δHBV-DA-PS | 0.82 | 0.852 | 0.01 | 2.66 |
| LSTM-DI (1) | 0.84 | 0.855 | 0.04 | 2.83 |
| LSTM | 0.74 | 0.743 | 0.05 | 3.54 |
| LSTM-DA | 0.82 | 0.862 | 0.01 | 2.63 |

Either precipitation or state adjustment can mitigate the underestimation of high flow by increasing precipitation or soil water storage (Table 1). The precipitation adjuster shows benefits in the western US while negatively impacting simulations in the central US and the southeastern coastal region (Figure 2a). It can increase the precipitation to help correct an otherwise underestimated peak flow response (e.g., March 1991 in Figure 3i), or adjust the timing of peaks (e.g., Apr 1992 in Figure 3i), but cannot address the storage-related water release processes. Thus, precipitation adjustment is effective when the background model already performs well in the recession limb, but the forcing product has high errors for the peaks. However, the precipitation adjustment can cause larger errors than the background model when correcting single peaks in "flashy" (short-term, high-flow) events in the dry season, due to the time lag of the adjuster. With such a short-term peak, the adjuster ends up being applied on a zero-precipitation day, making it ineffective (Figure 3ii, southeastern coastal US; —similar instances were observed for dry basins in the central US). In contrast, the advantages of the state adjuster are more evenly distributed across CONUS, with the exceptions of the southern Appalachian region and the Interior Highlands (Figure 2b). The state adjustor enhances peak flows by increasing upper soil storage (SUZ) for fast flows (Figure 3v), and controls the recession period by adjusting lower soil storage (SLZ) (Figure 3iii) associated with baseflow release. This type of DA can be effective when both peak and recession limbs need correction due to model structural errors or accumulated long-term forcing errors. However, the state adjuster cannot respond quickly for correction of short-term errors, as its effectiveness in adjusting states relies on modifying the states for the first day in the DA window rather than those of the last (most recent) day (Eq. 9 and 10). It also struggles with storm-induced streamflow that involves frequent peaks (Figure 3iv and vi). Nonetheless, as the precipitation and state adjusters both improve high flow predictions but work on different flow mechanisms, their benefits can be superimposed for greater overall effectiveness (Figure 3vii).

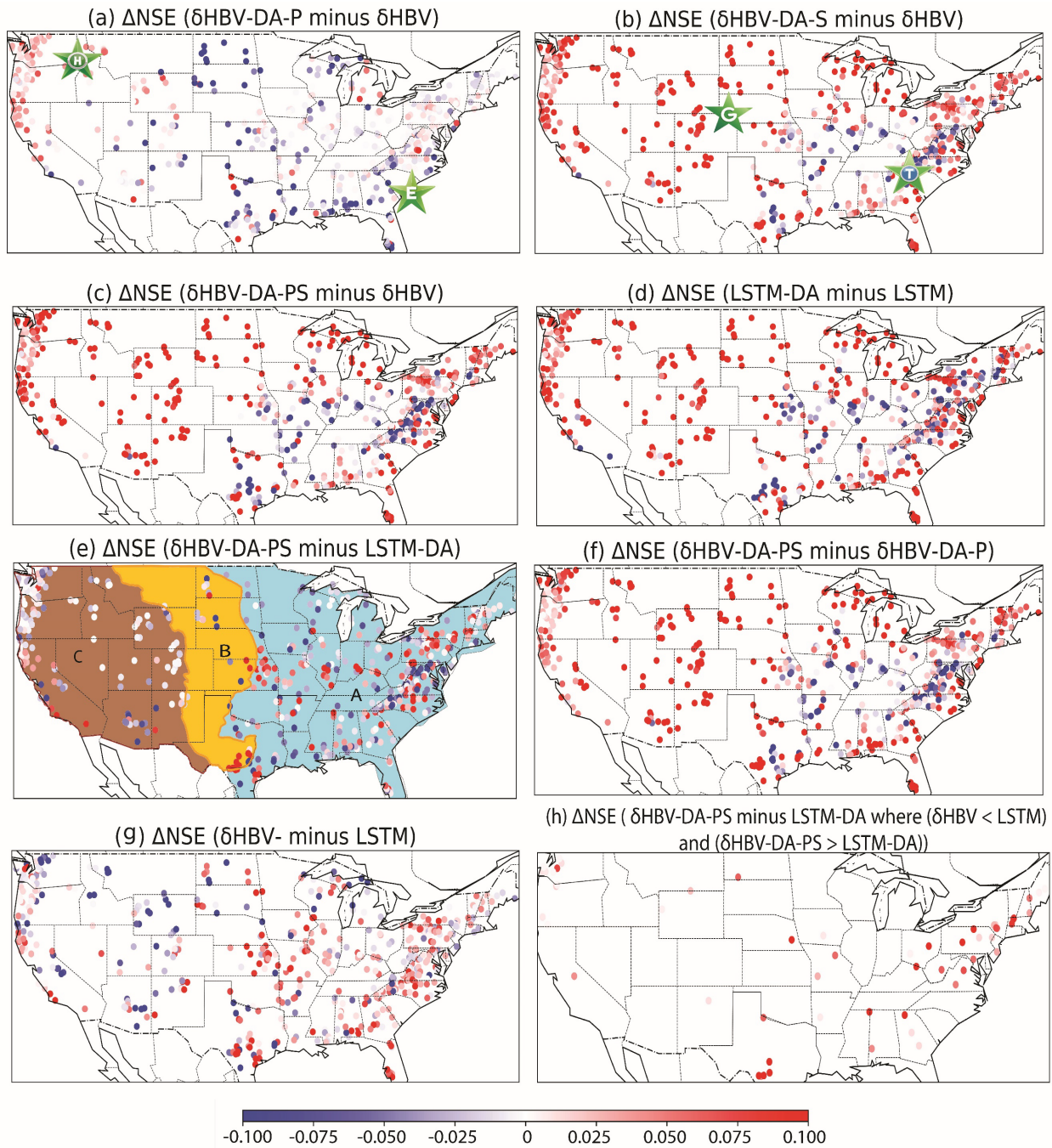

Figure 2: NSE difference maps between (a) $\delta$HBV-DA-P and $\delta$HBV, (b) $\delta$HBV-DA-S and $\delta$HBV, (c) $\delta$HBV-DA-PS and $\delta$HBV, (d) LSTM-DA and LSTM, (e) $\delta$HBV-DA-PS and LSTM-DA, (f) $\delta$HBV-DA-PS and $\delta$HBV-DA-P, and (g) $\delta$HBV and LSTM. In map e, we have divided the USA into three physiographic regions: Eastern (blue, A), Central (gold, B), and Western (brown, C).

The boundaries of the Great Plains serve as the dividing lines, with areas west of the Great Plains designated as Western and areas east considered Eastern. The sites in subplots a-b marked with green star-shaped points represent the locations for the hydrograph plots in Figure 3.

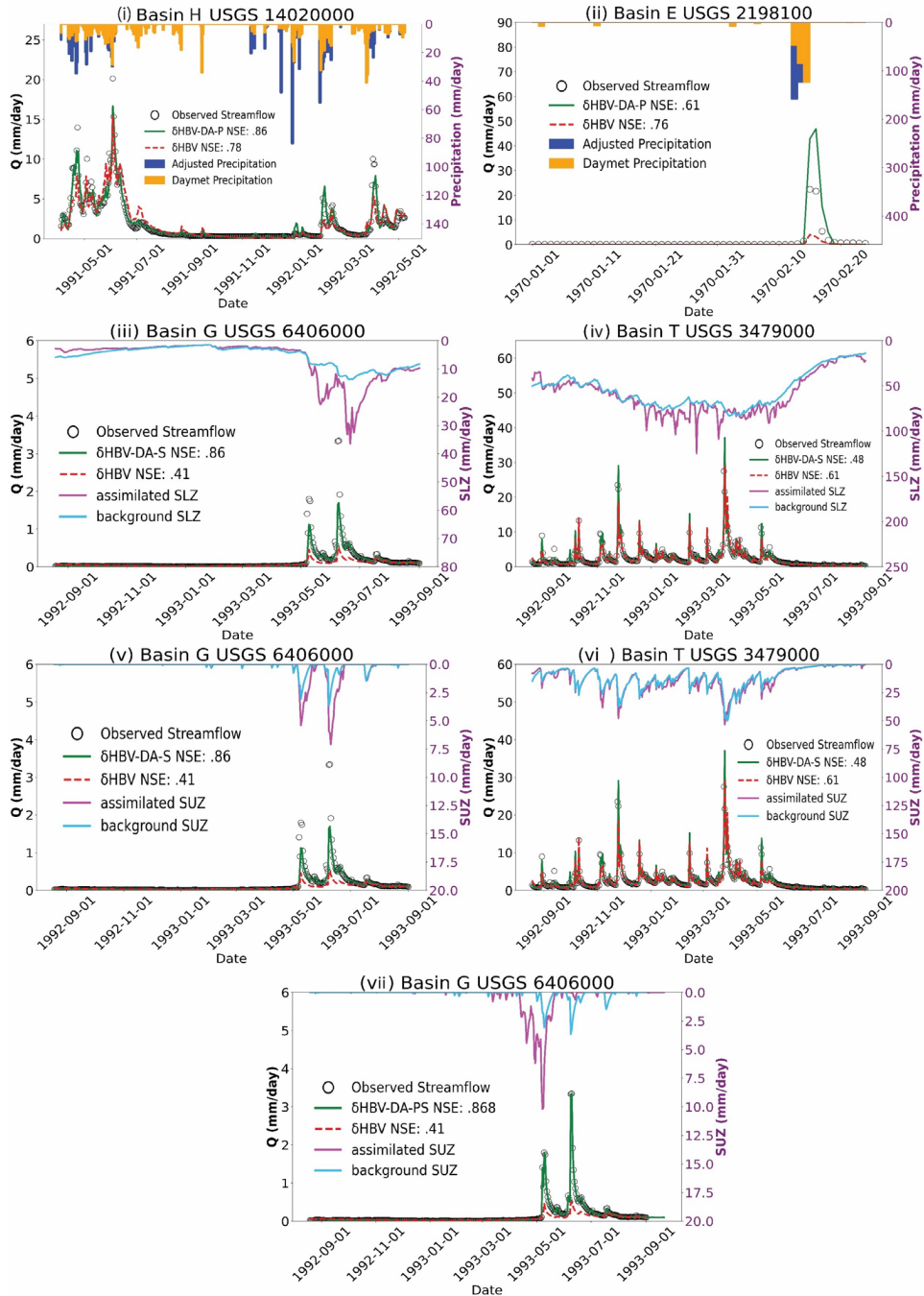

*Figure 3: Hydrographs of observations, original simulation, and streamflow after DA at USGS stations: (i) 14020000 UMATILLA RIVER ABOVE MEACHAM CREEK, OR; (ii) 2198100,*

*BEAVERDAM CREEK NEAR SARDIS, GA; (iii), (v), and (vii) 6406000, BATTLE CR AT HERMOSA SD; (iv) and (vi) USGS 3479000 WATAUGA RIVER NEAR SUGAR GROVE, NC. Subplots (i) and (ii) show the original Daymet and adjusted precipitation. (iii) to (vii) show the original and adjusted hydrological states from δHBV-DA-S and δHBV-DA-PS. NSE values of models across the whole testing period are listed in the subplots. The locations of all sites are marked by green stars in Figure 2.*

### 3.2. δHBV-DA vs. LSTM-DA

Both δHBV-DA-PS and LSTM-DA greatly enhanced streamflow simulation accuracy compared to the non-DA versions, achieving the same performance with a single source of forcing data (Daymet): an NSE value of 0.82. δHBV-DA-PS improved δHBV's median NSE by 0.07 (from 0.75 to 0.82), while LSTM-DA improved LSTM's NSE by 0.08 (from 0.74 to 0.82). LSTM-DI(1), being LSTM-DI with a one-day-lagged observation directly used as an input, presented a slightly higher NSE (0.84) than either δHBV-DA-PS or LSTM-DA but had higher RMSE values for both low and high flow prediction. This comparison is consistent with the results in Nearing et al. (2022) and Feng et al. (2020): DI (i.e., autoregression in Nearing et al. (2022)) is moderately more effective than DA. DA only updates states or specific input variables (e.g., precipitation) while DI for LSTM has more freedom in updating both states and weights in neural networks using the recent observations. However, DA can update the simulation whenever observations are available, without requiring historical data to train the model, which is particularly valuable for ungauged basins.

Overall, δHBV-DA-PS improved the simulation against δHBV in 80% of basins, while LSTM-DA improved 74% of basins against LSTM (Figure 3a&b). δHBV-DA-PS outperformed LSTM-DA for low-performing basins (NSE < 0.8), while LSTM-DA exhibited a slight advantages for basins with NSE values above 0.8 (Figure 3c). We divided the CONUS into three regions—Eastern (Region A), Central (Region B), and Western (Region C)—based on physiographic regions (Figure 2e). δHBV-DA-PS performed better than LSTM-DA mostly in the Eastern US (Table 3). It also showed a slight advantage in the Central region (Region B), mainly in the southern Great Plains. In the Western region (Region C), LSTM-DA performed better than δHBV-DA-PS, except for some basins in California, where both schemes reached median NSE values over 0.9 in this region. On the one hand, the regions where δHBV-DA-PS had advantages over LSTM-DA — Eastern, south Central, and California — are similar to those where δHBV was stronger than LSTM (similar patterns are shown in Figures 2e and 2g). As δHBV has a process-based model as its backbone, it may have structural advantages in simulating hydrological processes in these regions than purely data-driven LSTM. On the other hand, δHBV-DA-PS also performed better than LSTM-DA for other basins within these regions where δHBV alone initially performed worse than LSTM (Figure 2h). A plausible reason is that LSTM-DA, with adjustments to a larger number of states, is prone to over-tuning (overfitting), whereas δHBV-DA-PS, having fewer states as well as stronger physical constraints between these states, can be more effective in these basins. However, in regions where rainfall-runoff processes are more complex and less accurately represented by HBV, such as the coastal Eastern US and Southwestern US, LSTM-DA offered greater advantages due to its freedom to adjust a larger number of states, allowing for better adaptation to complex dynamic processes. Both δHBV-DA and LSTM-DA performed worse in the southeastern US, such as the Southern Great Plains, Southern Appalachia, and Central Lowlands, where flashy and frequent runoff patterns dominate, since variational DA has limitations in handling rapidly-changing dynamics.

*Table 2: Statistic metrics of δHBV-DA-PS and LSTM-DA for each physiographic region (Figure 2e). Bold font indicates the best performance in each region.*

| Region | Model | NSE | Annual relative peak errors (%) | Low flow (mm/day) | High flow (mm/day) |
|---|---|---|---|---|---|
| Eastern US (A) | LSTM-DA | 0.770 | 23.54 | 0.01 | 2.74 |
| | δHBV-DA-PS | **0.782** | **21.91** | 0.01 | 2.68 |
| Central US (B) | LSTM-DA | 0.614 | 39.20 | 0.01 | 1.26 |
| | δHBV-DA-PS | **0.660** | **33.28** | 0.01 | 1.05 |
| Western US (C) | LSTM-DA | **0.924** | **12.70** | 0.01 | **2.42** |
| | δHBV-DA-PS | 0.907 | 14.58 | 0.01 | 3.10 |

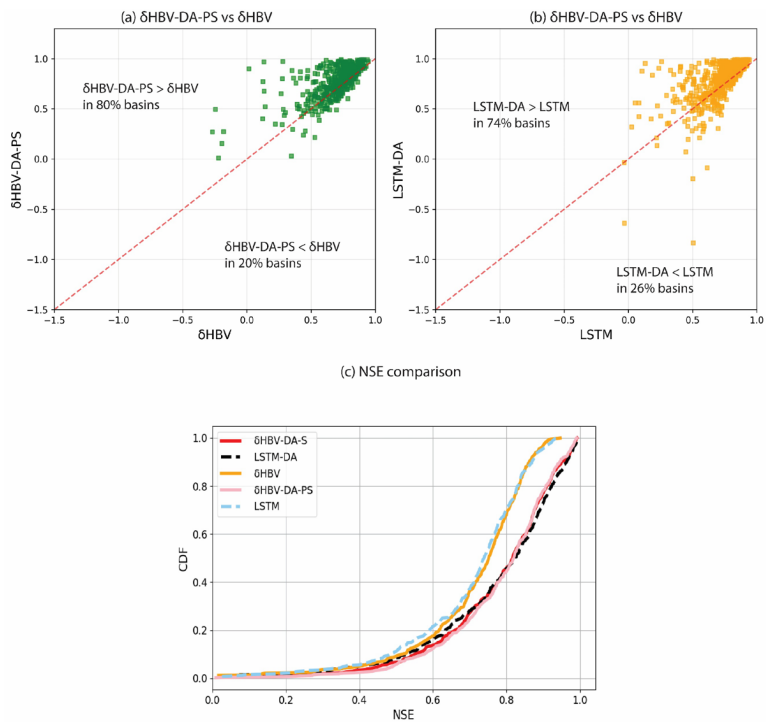

*Figure 4: (a) Scatter plot comparing basin NSE values between δHBV with data assimilation using (δHBV-DA-PS) and the baseline δHBV model. Each point represents one basin. This plot illustrates that δHBV-DA-PS performed better than δHBV in 80% of the basins, as shown by the majority of data points being above the dashed red line y = x. (b) Scatter plot comparing basin*

*NSE values between LSTM with data assimilation (LSTM-DA) and the baseline LSTM model. This plot shows that LSTM-DA outperformed the baseline LSTM in 74% of the basins. (c) Cumulative Density Function (CDF) plot of basin NSE scores for different models.*

### 3.3. Further discussion

Overall, differentiable models with limited physical states can benefit from variational DA just as much as LSTM, achieving comparable levels of accuracy across all metrics. Unlike LSTM—which functions as a black box—the HBV model (and its differentiable version, δHBV) offers full interpretability of rainfall-runoff processes. We observed that DA systematically adjusted physical states in alignment with flow mechanics, modifying the upper soil zone for faster flow and the lower soil zone for slower base flow. By coordinating the storage levels of these zones, the model effectively enhanced peak flow predictions and improved the subsequent recession limbs. Notably, embedding physical equations did not restrict the differentiable model's ability to benefit from data assimilation; rather, it illuminated the physical processes involved and helped pinpoint potential sources of error, guiding future model refinements.

Both precipitation and state adjusters can significantly improve peak flow predictions. When they operate simultaneously—such as in δHBV-DA-PS—their effects may compensate for one another or be superimposed. Although the variational data assimilation developed for δHBV with a few states has proven effective, it remains overly flexible, leading to an underdetermined problem. For example, discrepancies arise between states adjusted solely by the state adjuster and those adjusted when both components are applied (Figure 3v and vii). This method can be further improved to preserve the spatial pattern of simulation with more physical or data constraints. This work builds a benchmark for future comparison with multiple alternatives, such as Ensemble Kalman Filter and Particle Filter.

## 4. Conclusions

Our findings indicate that differentiable physics-informed machine learning models, when enhanced with data assimilation techniques—such as our δHBV-DA-PS model—can effectively compete with state-of-the-art "pure" deep learning models using data assimilation (LSTM-DA). Crucially, the δHBV-DA-PS model incorporates physically based hydrological state and precipitation adjusters, offering a structured approach to explicitly capture the natural dynamics of water movement and storage. The process representation also provides insights regarding the mechanisms of the corrections, which is important for interpretation.

Exploring other data assimilation approaches such as Particle Filtering and Kalman Filters should provide deeper insights into their accuracy and effectiveness in various hydrological scenarios. These techniques are renowned for their ability to manage uncertainties within model forecasts and could be key in refining the predictive accuracy of hydrological models across diverse environmental conditions.

**Acknowledgments**

This work was supported by subaward A23-0271-S001 from the Cooperative Institute for Research to Operations in Hydrology (CIROH) through the National Oceanic and Atmospheric Administration (NOAA) Cooperative Agreement NA22NWS4320003. The statements, findings, conclusions, and recommendations are those of the authors and do not necessarily reflect the view of NOAA. The authors gratefully acknowledge the computing resources—both CPU and GPU—provided by Pantarhei, an analytical and computational resource accessible to the CIROH research community. This cluster is managed by the CIROH IT Computing group at the University of Alabama.

**Data Availability**

All data used in this work are publicly available. The CAMELS dataset, which includes streamflow observations, geographic attributes, and the forcing dataset (Daymet) used in this work, can be downloaded at https://dx.doi.org/10.5065/D6MW2F4D (Addor et al., 2017; Newman & Clark, 2014). The differentiable model code used in this work was previously published (https://doi.org/10.5281/zenodo.7091334). The LSTM and LSTM-DA codes can be accessed at https://zenodo.org/records/7063252, and the LSTM-DI code can be found on GitHub (https://github.com/mhpi/hydroDL). The data assimilation code of δHBV will be released with the paper acceptance.

**Appendix**

*A. Model input data*

*Table A1. Summary of the forcing and attribute variables used in all δHBV models. All variables are from the CAMELS dataset (Addor et al., 2017; Newman & Clark, 2014)*

| Type | Variable | Description [units] | Variable | Description [units] |
|---|---|---|---|---|
| Forcings | PRCP | Precipitation [mm/day] | T | Temperature [°C] |
|  | $E_p$ | Potential evapotranspiration [mm/day] | - | - |
| Attributes | p_mean | Mean daily precipitation [mm/day] | dom_land_cover | Dominant land cover type [-] |
|  | pet_mean | Mean daily PET [mm/day] | root_depth_50 | Root depth at 50th percentile, extracted from a root depth distribution based on the International Geosphere-Biosphere Programme (IGBP) land cover [m] |
|  | p_seasonality | Seasonality and timing of precipitation [-] | soil_depth_pelletier | Depth to bedrock [-] |
|  | frac_snow | Fraction of precipitation | soil_depth_statgso | Soil depth [m] |
|  |  | falling as snow [-] |  |  |
|  | Aridity | PET/P [-] | soil_porosity | Volumetric soil porosity soil_conductivity [-] |
|  | high_prec_freq | Frequency of high precipitation days [days/year] | soil_conductivity | Saturated hydraulic conductivity [cm/hr] |
|  | high_prec_dur | Average duration of high precipitation events [days] | max_water_content | Maximum water content [m] |
|  | low_prec_freq | Frequency of dry days [days/year] | sand_frac | Sand fraction [-] |
|  | low_prec_dur | Average duration of dry periods [days] | silt_frac | Silt fraction [-] |
|  | elev_mean | Catchment mean elevation [m] | clay_frac | Clay fraction [-] |
|  | slope_mean | Catchment mean slope [m/km] | geol_class_1st | Most common geologic class in the catchment basin [-] |
|  | area_gages2 | Catchment area (GAGESII estimate) [km$^2$] | geol_class_1st_frac | Fraction of the catchment area associated with its most common geologic class [-] |
|  | frac_forest | Forest fraction [-] | geol_class_2nd | Second most common geologic class in the catchment basin [-] |
|  | lai_max | Maximum monthly mean of the leaf area index [-] | geol_class_2nd_frac | Fraction of the catchment area associated with its 2nd most common geologic class [-] |
|  | lai_diff | Difference between the maximum and minimum monthly mean of the leaf area index [-] | carbonate_rocks_frac | Fraction of the catchment area as carbonate sedimentary rocks [-] |
|  | gvf_max | Maximum monthly mean of the green vegetation [-] | geol_porosity | Subsurface porosity [-] |
|  | gvf_diff | Difference between the maximum and minimum monthly mean of the green vegetation fraction [-] | geol_permeability | Subsurface permeability [m$^2$] |

| | dom_land_cover_frac | Fraction of the catchment area associated with the dominant land cover [-] | - | - |
|---|---|---|---|---|

## B. Long short-term memory (LSTM)

LSTM has been thoroughly described in other work, so we only briefly discuss its structure here; more details on the LSTM implemented in this work can be found in Kratzert et al. (2019) and Nearing et al. (2022). In short, the LSTM architecture uses a cell state ($c$) to store long-term information and hidden states ($h$) to pass information between time steps. The cell state enables the LSTM to learn long-term dependencies. Input gate weights determine what enters the cell and its hidden states, forget gate weights remove information from the cell state, and output gate weights control the hidden states passed to the next step (Hochreiter & Schmidhuber, 1997). The LSTM streamflow prediction model can be summarized by the following equations:

Input gate: $$i^t = \sigma(W_{ix}x^t + W_{ih}h^{t-1} + b_i)$$ (B1.1)

Forget gate: $$f^t = \sigma(W_{fx}x^t + W_{fh}h^{t-1} + b_f)$$ (B1.2)

Input node: $$g^t = tanh(W_{gx}x^t + W_{gh}h^{t-1} + b_g)$$ (B1.3)

Output gate: $$o^t = \sigma(W_{ox}x^t + W_{oh}h^{t-1} + b_o)$$ (B1.4)

Cell state: $$s^t = g^t \odot i^t + s^{t-1} \odot f^t$$ (B1.5)

Hidden state: $$h^t = tanh(s^t) \odot o^t$$ (B1.6)

where the input, $x^t$, is a concatenation of meteorological forcings and static attributes at time step $t$. $ReLU$, $tanh$, and $\sigma$ represent the Rectified Linear Unit, hyperbolic tangent, and sigmoid activation functions, respectively. $W$ and $b$ are the neuron weights and bias for each layer (distinguished using subscripts). $\odot$ is element-wise multiplication. $h$ and $s$ represent the hidden states and cell states, which respectively carry short-term and long-term memory.

## C. Differentiable Hydrologiska Byråns Vattenbalansavdelning models

Table C1 provides the details of HBV, including its equations, fluxes, and parameters. In δHBV, three parameters ($\gamma$, $\beta$, and $K_0$) were set as dynamic (changing with time) to enhance the model's accuracy and responsiveness to environmental changes. Parameter $\gamma$ adapts to seasonal vegetation changes and drought recovery, $\beta$ accounts for past rainfall and soil moisture effects on runoff production, and $K_0$ adjusts to represent rapid water release rates from the upper layer.

*Table C1. δHBV Equations*

| Modules | Equations | Fluxes | Parameter |
|---|---|---|---|
| Snow accumulation and melt | $\frac{dS_p}{dt} = P_s + R_{fz} - S_{smelt}$ <br> $P_s = P$ if $T < \theta_{TT}$, otherwise 0 <br> $R_{fz} = (\theta_{TT} - T)\theta_{DD}\theta_{rfz}$ <br> $S_{melt} = (T - \theta_{TT})\theta_{DD}$ <br> $\frac{dS_{liq}}{dt} = s_{melt} - R_{fz} - I_{snow}$ <br> $I_{snow} = s_{liq} - \theta_{CWH}S_p$ | $S_p$: snow storage <br> $P_s$: precipitation as snow <br> $R_{fz}$: refreezing of liquid snow <br> $s_{melt}$: snowmelt as water equivalent <br> $S_{liq}$: liquid water content in the snowpack <br> $I_{snow}$: snowmelt infiltration to soil moisture | $\theta_{TT}$: threshold temperature for snowfall [°C] <br> $\theta_{DD}$: degree-day factor [mm°C$^{-1}$day$^{-1}$] <br> $\theta_{rfz}$: refreezing coefficient [-] <br> $\theta_{CWH}$: water holding capacity as a fraction of the current snowpack [-] |
| Soil moisture and evapotranspiration | $\frac{dS_S}{dt} = I_{snow} + P_r - P_{eff} - E_x - E_T(+C_r)$ <br> $P_r = P$ if $T > \theta_{TT}$ otherwise 0 <br> $P_{eff} = \min\left(\left(\frac{S_S}{\theta_{FC}}\right)^\beta, 1\right)(P_r + I_{snow})$ <br> $E_x = (S_S - \theta_{FC})/dt$ <br> $E_T = \min\left(\left(\frac{S_S}{\theta_{FC}\theta_{LP}}\right)^\gamma, 1\right)E_P$ <br> $C_r = \theta_C * S_{LZ} * \left(1 - \frac{S_S}{\theta_{FC}}\right)$ | $S_S$: storage in soil moisture <br> $S_{LZ}$: current storage in the lower subsurface zone <br> $P_r$: precipitation as rain <br> $P_{eff}$: effective flow to the upper subsurface zone <br> $E_x$: rainfall excess <br> $E_T$: actual evapotranspiration. <br> $C_r$: an upward flow from the lower subsurface zone. This flux is only used in δHBV1.1p | $\theta_C$: time parameter [day$^{-1}$] <br> $\theta_{FC}$: maximum soil moisture (field capacity) [mm] <br> $\theta_{LP}$: vegetation wilting point [-] <br> $\beta$: a parameter influencing the shape of the soil moisture function [-] <br> $\gamma$: a parameter influencing the shape of the evapotranspiration function [-] |
| Runoff generation | $\frac{dS_{UZ}}{dt} = P_{eff} + E_x - perc - Q_0 - Q_1$ <br> $Perc = \min(\theta_{perc}, S_{UZ}/dt)$ <br> $Q_0 = \theta_{K_0}(S_{UZ} - \theta_{UZL})$ <br> $Q_1 = \theta_{K_1}S_{UZ}$ <br> $\frac{dS_{LZ}}{dt} = Perc - Q_2(-C_r)$ <br> $Q_2 = \theta_{K_2}S_{LZ}$ <br> $Q' = Q_0 + Q_1 + Q_2$ <br> $Q_b(t) = \int_0^t \xi(s)Q'(t-s)ds$ <br> $\xi(s) = \frac{1}{\Gamma(\theta_a)\theta_b^{\theta_a}}s^{\theta_a-1}e^{-\frac{1}{\theta_b}}$ | $S_{UZ}$: storage in the upper subsurface zone <br> $perc$: percolation to the lower subsurface zone <br> $Q_0$: near surface flow <br> $Q_1$: interflow <br> $Q_2$: baseflow <br> $S_{LZ}$: storage in the lower subsurface zone <br> $Q_b$: simulated streamflow | $\theta_{perc}$: percolation flow rate [mm/day] <br> $\theta_{K_0}, \theta_{K_1},$ and $\theta_{K_2}$: recession coefficients [day$^{-1}$] <br> $\theta_a$ and $\theta_b$: two routing parameters [-] <br> $\theta_{UZL}$: maximum storage of the upper soil layer [mm] |